\documentclass[12pt]{article}
\usepackage{amssymb}
\usepackage{epsf}
\newcommand{\simgt}{\lower.5ex\hbox{$\; \buildrel > \over \sim \;$}}
\newcommand{\simlt}{\lower.5ex\hbox{$\; \buildrel < \over \sim \;$}}
\begin{document}

\markboth{K.Takahashi, K. Sato, H. E. Dalhed and J. R. Wilson}
{Shock propagation and neutrino oscillation}

\title{
Shock propagation and neutrino oscillation in supernova
}

\author{
K. Takahashi$^*$, K. Sato$^{*,**}$,
H. E. Dalhed$^{***}$ and J. R. Wilson$^{***}$
\\
$^*$Department of Physics, University of Tokyo,
\\
7-3-1 Hongo, Bukyo, Tokyo 113-0033, Japan
\\
$^{**}$Research Center for the Early Universe, University of Tokyo, 
\\
7-3-1 Hongo, Bukyo, Tokyo 113-0033, Japan
\\
$^{***}$Lawrence Livermore National Laboratory,
\\
7000 East Avenue, L-015, Livermore, CA 94550}

\maketitle

\baselineskip=16pt

\begin{abstract}

The effect of the shock propagation on neutrino oscillation in supernova 
is studied paying attention to evolution of average energy of $\nu_{e}$ 
and $\bar{\nu}_{e}$. We show that the effect appears as a decrease in
average $\nu_{e}$ (in case of inverted mass hierarchy, $\bar{\nu}_{e}$) 
energy at stellar surface as the shock propagates. It is found that
the effect is significant 2 seconds after bounce if 
$3 \times 10^{-5} < \sin^{2}{\theta_{13}} < 10^{-2}$.

\end{abstract}

\section{Introduction}

Recently effects of shock propagation on neutrino oscillation in supernova
was studied \cite{SchiratoFuller2002,Lunardini2003} and it was shown that some
characteristic signatures emerge as the shock propagates through
the regions where matter-enhanced neutrino flavor conversion occurs.

There have been many studies on neutrino oscillation in supernova:
extracting information of neutrino parameters from the observation of
SN1987A neutrinos 
\cite{Smirnov1994,Jegerlehner1996,MinakataNunokawa2001,Lunardini2001} or
a future supernova neutrinos \cite{Dighe2000,KT1,KT4,KT2,KT3},
and probing supernova physics from observed neutrinos of a future supernova 
\cite{Minakata2001}. But all of them are done without the effect of 
the shock propagation.

In this paper the effect of the shock propagation is studied paying attention
to evolution of average energies of $\nu_{e}$ and $\bar{\nu}_{e}$.
We show when and with which parameter ($\sin^{2}{2\theta_{13}}$) the 
effect is significant or can be neglected safely.

\section{Neutrino oscillation and shock propagation}

If mixing angle is small, dynamics of flavor conversions is
well described by resonant oscillation, which occurs at density,
\begin{equation}
\rho_{\rm res} \simeq 1.4 \times 10^{6} {\rm g/cc}
\left( \frac{\Delta m^{2}}{1 {\rm eV}^{2}} \right)
\left( \frac{10 {\rm MeV}}{E_{\nu}} \right)
\left( \frac{0.5}{Y_{e}} \right)
\cos{2 \theta},
\end{equation}
where $\Delta m^{2}$ is the mass squared difference, $\theta$ is the
mixing angle, $E_{\nu}$ is the neutrino energy, and $Y_{e}$ is the mean 
number of electrons per baryon. Flavor conversion probabilities are 
determined by adiabaticity parameter $\gamma$:
\begin{equation}
\gamma \equiv \frac{\Delta m^{2}}{2 E_{\nu}} 
\frac{\sin^{2}{2 \theta}}{\cos{2 \theta}} 
\frac{n_{e}}{\left|dn_{e}/dr\right|}.
\label{eq:adiabaticity}
\end{equation}
Here $\Delta m^{2}$ and $\theta$ are
\begin{eqnarray}
\theta_{13} \; & {\rm and} & \; \Delta m^{2}_{13} \; {\rm at \; H-resonance},
\nonumber \\
\theta_{12} \; & {\rm and} & \; \Delta m^{2}_{12} \; {\rm at \; L-resonance},
\nonumber
\end{eqnarray}
where mixing matrix is taken as:
\begin{equation}
U  =  \left(\begin{array}{ccc}
c_{12}c_{13} & s_{12}c_{13} & s_{13}\\
-s_{12}c_{23}-c_{12}s_{23}s_{13} & c_{12}c_{23}-s_{12}s_{23}s_{13} 
& s_{23}c_{13}\\
s_{12}s_{23}-c_{12}c_{23}s_{13} & -c_{12}s_{23}-s_{12}c_{23}s_{13} 
& c_{23}c_{13}
\end{array}\right)\label{mixing_matrix},
\end{equation}
where $s_{ij} = \sin{\theta_{ij}}, c_{ij} = \cos{\theta_{ij}}$ 
for $i,j=1,2,3 (i<j)$.
When $\gamma \ll 1$, the resonance is nonadiabatic and 
the fluxes of the two involved mass eigenstates are completely exchanged. 
On the contrary, when $\gamma \gg 1$, the resonance is adiabatic and the 
conversion between mass eigenstates does not occur. 

\begin{figure}[t]
\epsfxsize=10cm
\centerline{\epsfbox{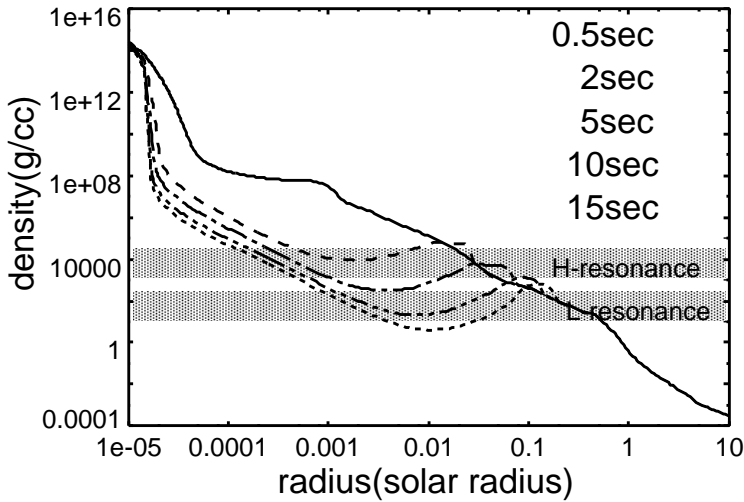}}
\caption{Evolution of density profile after bounce. Densities which 
correspond to H- and L-resonance are also shown.}
\label{fig:density}
\end{figure}

Evolution of density profile is shown in Fig. \ref{fig:density}.
This is calculated by our numerical supernova model with progenitor mass,
$18 M_{\odot}$. (For detail, see \cite{Wilson1986}.)
As the shock passes through the matter outside the neutrinosphere, 
densities will typically increase by approximately a factor of two.  
As neutrinos subsequently heat the matter behind the shock,
a high entropy region will develop which accelerates matter behind the shock,
causing a density buildup between this high entropy bubble and the outgoing
shock. The density gradient caused by this mechanism is typically several
orders of magnitude greater than that caused by the shock passage.  
As seen in Fig. \ref{fig:density}, the shock is barely visible on 
the outer part of the density gradient at approximately 0.1 solar radius. 
This "snowplowed" density gradient is typical of neutrino driven core 
collapse supernovae, and is expected to be generic of a range of 
progenitor masses.

Densities which correspond to H- and L-resonance are
also shown (Note that the resonance density depends on neutrino energy).
As can be seen, the shock reaches the resonant region about 2 seconds
after bounce and the density behind the shock is lower than that forward 
the shock. In the late phase ($t >$ several seconds), a neutrino experience 
three times of H-resonances (and/or L-resonances) and the radii of resonance 
points ($r_{\rm res}$) are in general much smaller than the early phase. Since 
$\gamma \propto r_{\rm res}$ if the density profile is approximated to be
power-law, in general the three resonances have different adiabaticity 
paremeters which are much smaller than that of the early phase. 
Therefore the average energy of the observed neutrinos
is expected to depend on time not only due to the evolution of
the neutrinosphere and the protoneutron star but also due to the
evolution of the shock. 

We can estimate the above effect by calculating adiabaticity parameter. 
We consider the case where H-resonance occurs only one time and assume that 
the neutrino mass hierarchy is normal. In this case the 
adiabaticity parameter is, assuming $\rho \propto r^{-n}$,
\begin{equation}
\gamma_{\rm H} \approx 2 \times 10^{2} n^{-1}
\left(\frac{\sin^{2}{2\theta_{13}}}{10^{-2}}\right)
\left(\frac{r_{\rm res}}{3 \times 10^{-2} R_{\odot}}\right)
\left(\frac{\Delta m^{2}}{3 \times 10^{-3}}\right)
\left(\frac{10 {\rm MeV}}{E_{\nu}}\right).
\end{equation}
Note that the index $n$ is almost independent of time as can be seen
in Fig. \ref{fig:density}. From this H-resonace is expected to be adiabatic if 
$\sin^{2}{2\theta_{13}} > 10^{-4}$. Adiabaticity parameter, however, becomes 
smaller as the shock propagates and two order smaller at 15 sec than in the 
early phase. Thus the H-resonance becomes less adiabatic unless 
$\sin^{2}{2\theta_{13}} > 10^{-2}$ as is in the model LMA-L 
($\sin^{2}{2\theta_{13}} = 0.043$) of \cite{KT1}.
This will cause decreasing of $\nu_{e}$ average energy at the stellar surface 
as the shock propagates. On the other hand, if H-resonance is non-adiabatic 
even in early phase as is in the model LMA-S 
($\sin^{2}{2\theta_{13}} = 10^{-6}$) of \cite{KT1}, average energy 
will not change. Here it should be noted that almost half the neutrinos
are emitted after 2 seconds after bounce \cite{Totani1998}.

With inverted mass hierarchy, H-resonance occurs at anti-neutrino sector.
In this case evolution of average energy at stellar surface is seen in 
$\bar{\nu}_{e}$. Its qualitative feature is expected to be the same
as that of $\nu_{e}$ with normal mass hierarchy because it is determined
by the behavior of the shock.

As to L-resonance, the adiabaticity parameter is, substituting the
LMA (Large Mixing Angle) solution of the solar neutrino problem 
\cite{Fukuda2001,SNO,SNO2,KamLAND},
\begin{equation}
\gamma_{\rm L} \approx 4 \times 10^{3} n^{-1}
\left(\frac{\sin^{2}{2\theta_{12}}}{1}\right)
\left(\frac{r_{\rm res}}{0.3 R_{\odot}}\right)
\left(\frac{\Delta m^{2}}{7 \times 10^{-5}}\right)
\left(\frac{10 {\rm MeV}}{E_{\nu}}\right).
\end{equation}
Thus L-resonance will remain to be adiabatic even if the radius of the 
resonance point becomes two order smaller than the early phase.

In the next section we study the above effect quantitatively by numerical
calculation and obtain time evolution of the average energies of observed 
neutrinos ($\nu_{e}$ and $\bar{\nu}_{e}$).

\section{Numerical calculation}

We solve numerically evolution equations of neutrino wave functions along
the density profiles shown in Fig. \ref{fig:density}. From the wave
functions, we obtain flavor conversion probabilities, from which neutrino
spectra can be obtained by multiplying by the original neutrino fluxes.
To make the shock propagation effect distinctive, the original energy spectra 
at each time are set to be the same as the time-integrated spectra.
Neutrino parameters are taken as:
\begin{eqnarray}
&& \sin^{2}{2\theta_{12}} = 0.87, \;\;\;\; \sin^{2}{2\theta_{23}} = 1, 
\nonumber \\
&& \Delta m^{2}_{12} = 7.0 \times 10^{-5} {\rm eV^{2}}, \;\;\;\;
\Delta m^{2}_{13} = 3.2 \times 10^{-3} {\rm eV^{2}}.
\end{eqnarray}
As for $\sin^{2}{2\theta_{13}}$, we take various values including
values corresponding to model LMA-L and LMA-S in \cite{KT1,KT4}.
For detail of the calculational method and the original neutrino fluxes, 
see \cite{KT1,KT4}.

\begin{figure}[t]
\begin{center}
\epsfxsize=7.5cm
\epsffile{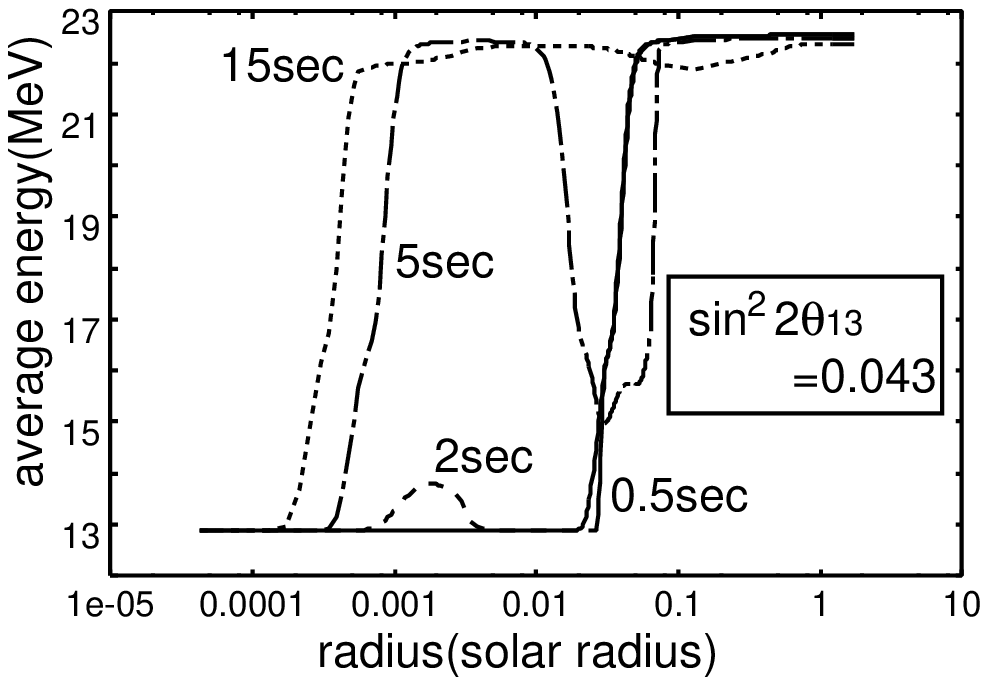}
\epsfxsize=7.5cm
\epsffile{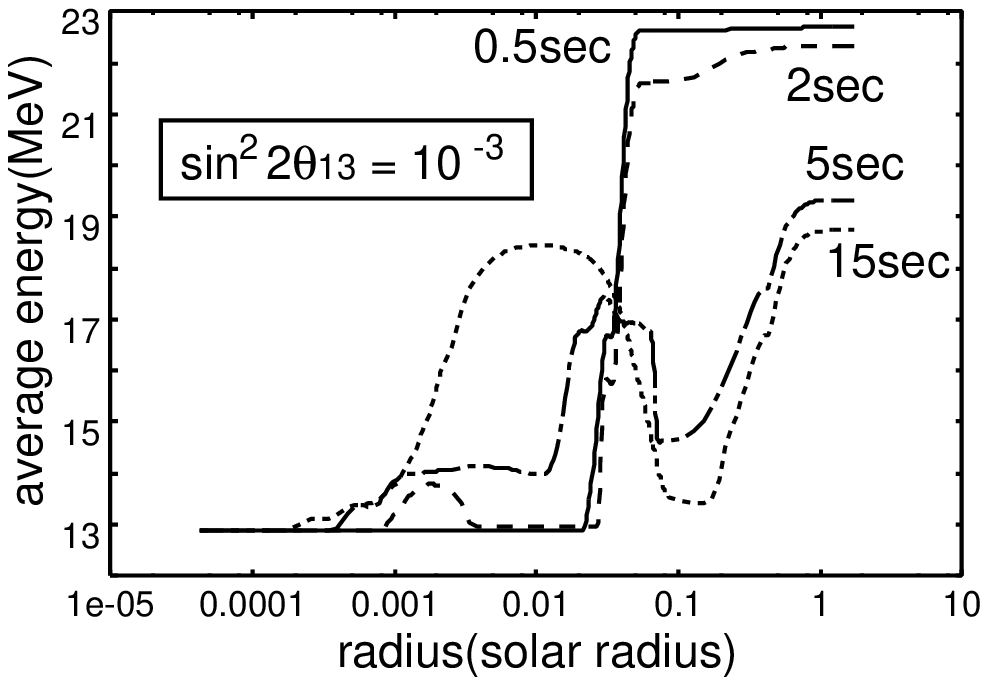}
\epsfxsize=7.5cm
\epsffile{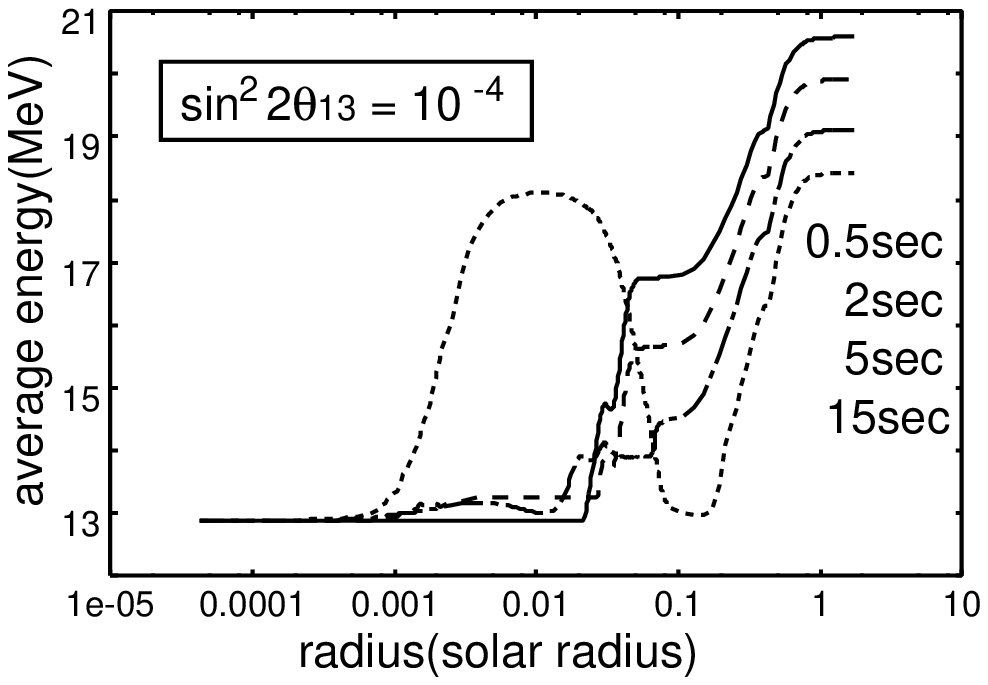}
\epsfxsize=7.5cm
\epsffile{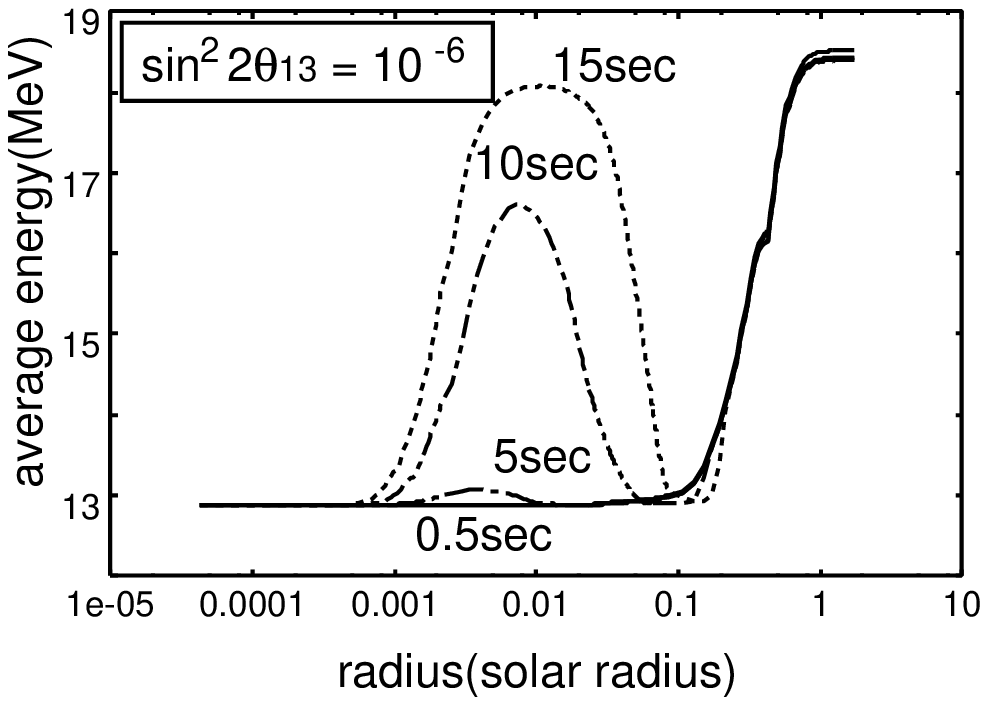}
\caption{Average energy evolutions of $\nu_{e}$ emitted at various times 
after bounce with normal mass hierarchy. The values of $\sin^{2}{2\theta_{13}}$
are that of the model LMA-L of \cite{KT1,KT4}, $10^{-3}$, $10^{-4}$ and
that of the model LMA-L of \cite{KT1,KT4}, respectively .
\label{fig:ave_ene}}
\end{center}
\end{figure}

Fig. \ref{fig:ave_ene} show average energy evolutions
of $\nu_{e}$ emitted at various times after bounce with normal mass 
hierarchy. Each figure differs in the value of $\sin^{2}{2\theta_{13}}$. 
The interesting behavior of the average energy in supernova (for example,
5 sec of the upper-left of Fig. \ref{fig:ave_ene}) indicates that H-resonance
occurs three times. As is discussed in the previous section, the average 
energy at the stellar surface does not change in time when 
$\sin^{2}{2\theta_{13}}$ is enough large (LMA-L, the upper-left of 
Fig. \ref{fig:ave_ene}) or small (LMA-S, the lower-right of 
Fig. \ref{fig:ave_ene}). On the other hand, average energy decrease by 
several MeV in the intermediate cases. By calculating with various values of 
$\sin^{2}{2\theta_{13}}$, we find that the shock propagation effects can be
seen when $3 \times 10^{-5} < \sin^{2}{2\theta_{13}} < 10^{-2}$ but
are absent till $\sim 1$ second after bounce irrespective of 
$\sin^{2}{2\theta_{13}}$.

\begin{figure}[t]
\epsfxsize=9cm
\centerline{\epsfbox{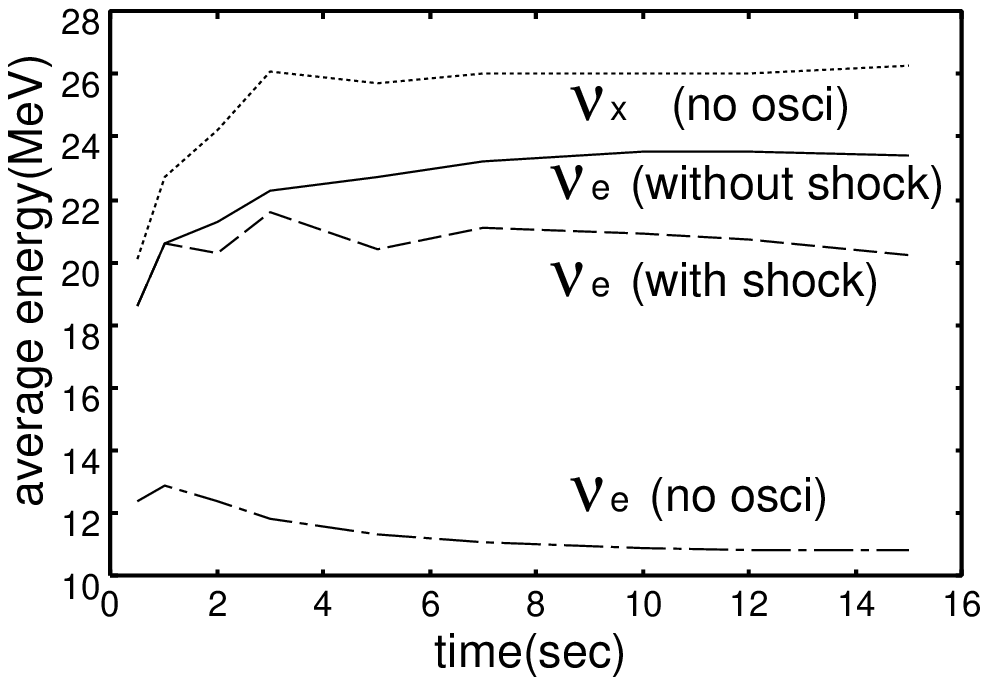}}
\caption{Evolutions of average energy of observed neutrinos.
Average energies of $\nu_{e}$ with and without shock effect are shown.
Those of $\nu_{e}$ and $\nu_{x} = \nu_{\mu},\nu_{\tau}$ without
neutrino oscillation are also shown. $\sin^{2}{2\theta_{13}}$ is set 
to $10^{-4}$.}
\label{fig:ave_ene_evo}
\end{figure}

Thus average energy of observed $\nu_{e}$ will change in time due to
shock propagating effect. In fact neutrino average energies changes
also due to evolution of protoneutron star and neutrinosphere.
Fig. \ref{fig:ave_ene_evo} shows evolutions of average energy of observed
neutrinos taking intrinsic changes of neutrino average energies
into account. $\sin^{2}{2\theta_{13}}$ is set to $10^{-4}$. As can also
be seen in the lower-left of Fig. \ref{fig:ave_ene}, $\nu_{e}$ energy with 
shock effect is lower by several MeV after about 2 sec after bounce than 
without shock effect.

As stated in the previous section, in case of inverted mass hierarchy 
it is $\bar{\nu}_{e}$ that is affected by shock propagation. Features of the 
evolution of the average energy are almost the same quantitatively: values of 
$\sin^{2}{2\theta_{13}}$ and time after bounce, with which shock propagation 
effect is significant, difference between average energies of the early phase 
and late phase.

\section{Discussion and summary}

As we saw in the previous section, neutrino average energy decrease in
general as the shock propagates. This is because the shock propagation
cause decrease in the adiabaticity parameter of H-resonance and suppress the
conversion between flavors. But it should be noted that the original spectra 
will change due to the evolution of the protoneutron star and the 
neutrinosphere. Thus we can not say about the value of $\sin^{2}{2\theta_{13}}$
and the mass hierarchy only from the evolution of the average energy of the
observed neutrino. To do so, we need a model of supernova and
spectrum evolution.

In this paper we studied the effect of the shock propagation paying attention
to evolution of average energies of $\nu_{e}$ and $\bar{\nu}_{e}$.
It is shown that the effect appears as a decrease in
average $\nu_{e}$ (in case of inverted mass hierarchy, $\bar{\nu}_{e}$) 
energy at stellar surface as the shock propagates. Further it is found that
the effect is significant 2 seconds after bounce if 
$3 \times 10^{-5} < \sin^{2}{\theta_{13}} < 10^{-2}$.

\section*{Acknowledgements}

K.T.'s work is supported by Grant-in-Aid for JSPS Fellows.
K.S.'s work is supported by Grant-in-Aid for Scientific Research (S) No.
14102004 and Grant-in-Aid for Scientific Research on Priority Areas No.
14079202.
H.E.D. and J.R.W. are supported by LLNL contract W-7405-ENG-48.

\end{document}